%% file: main.tex
\documentclass[conference,9pt]{IEEEtran}
\IEEEoverridecommandlockouts
\usepackage{cite}
\usepackage{amsmath,amssymb,amsfonts}
\usepackage{textcomp}
\def\BibTeX{{\rm B\kern-.05em{\sc i\kern-.025em b}\kern-.08em
    T\kern-.1667em\lower.7ex\hbox{E}\kern-.125emX}}
\ifCLASSINFOpdf
   \usepackage[pdftex]{graphicx}
\else
\fi
\usepackage{cite}
\usepackage{amsmath}
\usepackage{array}
\usepackage{lineno,hyperref}

\usepackage[usenames, dvipsnames]{color}
\usepackage[table]{xcolor}
\usepackage[linesnumbered,ruled,vlined]{algorithm2e}

\usepackage{algorithmicx}
\usepackage{algpseudocode}
\usepackage{amssymb}
\usepackage{xr}
\usepackage{multirow}
\usepackage{hhline}
\usepackage{array}
\usepackage{fancyvrb} 
\usepackage{multirow}

\usepackage[normalem]{ulem}

\MakeRobust{\Call}

\usepackage{amsmath}               
{
	
}
\usepackage[normalem]{ulem}
\graphicspath{{pics/}}

\usepackage{threeparttable}
\usepackage{dblfloatfix}    
\usepackage{comment}
\newtheorem{theorem}{\bf Theorem}

\usepackage{enumitem}
\usepackage{pifont}
\usepackage{todonotes}
\renewcommand{\arraystretch}{0.2}
\newcommand\T{\rule{0pt}{2ex}}       
\newcommand\B{\rule[-2.4ex]{0pt}{0pt}} 
\usepackage{flushend}
\usepackage{balance}

\begin{document}

\title{Cost-effective Design of Free-Flowing Microfluidic Networks
for Implementing Farey-Sequence Arithmetic Based\\
Sample Preparation Protocols
\title{Sample Preparation Meets Farey Sequence: A New Design Technique for
Free-Flowing Microfluidic Networks
}
}

\author{\IEEEauthorblockN{Tapalina Banerjee\IEEEauthorrefmark{1},
		Sudip Poddar\IEEEauthorrefmark{2},
		Tsung-Yi Ho\IEEEauthorrefmark{3}, and
		Bhargab B. Bhattacharya\IEEEauthorrefmark{4}}
	\IEEEauthorblockA{\IEEEauthorrefmark{1}Department of Computer Science and Engineering, JIS University, Kolkata, India}
	\IEEEauthorblockA{\IEEEauthorrefmark{2}Institute for Integrated Circuits, Johannes Kepler University Linz, Austria}
	\IEEEauthorblockA{\IEEEauthorrefmark{3}Department of Computer Science, National Tsing Hua University, Hsinchu, Taiwan}
	\IEEEauthorblockA{\IEEEauthorrefmark{4}Department of Computer Science and Engineering, Indian Institute of Technology Kharagpur, India}
	\IEEEauthorblockA{\{tapalinabanerjee1, sudippoddar2006, bhargab.bhatta\}@gmail.com, and tyho@cs.nthu.edu.tw}
	}	

\maketitle

\begin{abstract}
Design of microfluidic biochips has led to newer challenges
to the EDA community due to the availability of various flow-based
architectures and the need for catering to diverse applications such
as sample preparation, personalized medicine, point-of-care
diagnostics, and drug design. The ongoing Covid-19 pandemic has
increased the demand for low-cost diagnostic lab-on-chips manifold.
Sample preparation (dilution or mixing of biochemical fluids) is an
indispensable step of any biochemical experiment including sensitive
detection and successful assay execution downstream. Although for valve-based microfluidic biochips various design
automation tools are currently available, they are expensive, and prone to
various manufacturing and operational defects. Additionally, many problems
are left open in the domain of {\it free-flowing} biochips, where only a single
layer of flow-channels is used for fluid-flow devoid of any kind of control
layer/valves. In this work, we present a methodology for designing a free-flowing
biochip that is capable of performing fluid dilution according to
user’s requirement. The proposed algorithm for sample preparation
utilizes the Farey-sequence arithmetic of fractions that are used to
represent the concentration factor of the target fluid. We also
present the detailed layout design of a free-flowing microfluidic
architecture that emulates the dilution algorithm. The network
 is simulated using COMSOL multi-physics software accounting for relevant hydrodynamic parameters. Experiments on
various test-cases support the efficacy of the proposed design in
terms of accuracy, convergence time, reactant cost, and simplicity of
the fluidic network compared to prior art.

\end{abstract}


\IEEEpeerreviewmaketitle

\section{Introduction}
\label{Intro}
\input{Text/intro.tex}
\section{Preliminaries of sample preparation}
\label{DoP}
\input{Text/DoP}
\section{Related work}
\label{rworks}
\input{Text/related_work}
\section{Motivation}
\label{motiv}
\input{Text/motivation.tex}

\section{Proposed methodology}
\label{method}
\input{Text/Proposed_method.tex}

\section{Layout Design}
\label{lab:pysical_design}

\label{AlgoLayout}
\input{Text/AlgoLayout.tex}
\section{Experimental Results}
\label{ExpRes}
\input{Text/ExpRes}
\section{Conclusions}
\label{Concl}
\input{Text/Conclusions.tex}
\flushend
\bibliographystyle{./IEEEtran}
\bibliography{BibFile_FINAL}

\end{document}

%% file: Text/intro.tex
Microfluidic lab-on-chips (LoCs) have fueled the automation of biochemical protocols
on a tiny device by enabling the movement of fluids on a minuscule scale through microchannels as in Continuous-Flow Microfluidic Biochips (CFMBs), or on patterned-array of electrodes as discrete droplets for Digital Microfluidic Biochips (DMFBs).  An LoC ({\em a.k.a.}  biochip) encapsulates several fluidic modules that are capable of performing the desired sequence of fluidic functions needed to execute a biochemical protocol (assay), and sensors that report the outcome. LoCs provide fast and low-cost solutions to various applications such as sample-preparation, pathological diagnostics, DNA-sequencing, polymerase chain reaction (PCR), cell sorting, drug design, therapeutics, and synthetic biology, to name a few. The demand for biochips as a diagnostic kit has skyrocketed with the ongoing Covid-19 pandemic.

Continuous-flow microfluidic biochips ({CFMB}s) manipulate fluids in a
network of microchannels using pressure-driven microvalves. CFMBs 
can be of two types: (i) valve-based and (ii) valve-free (free-flowing).
The former type uses multi-layer technology to
fabricate flow-layer and control-layer. Micro-valves are actuated at the intersection of the
control and the flow layer to enable or stop fluid-flow through the latter. Many complex fluidic modules such as mixer, multiplexer, separator, pump, and storage units can be built on a single
chip utilizing thousands of micro-channels and valves \cite{LoC2010}. 
Although the use of control-valves provides the benefit of programmability,
these chips suffers from various fluidic errors due to various defects in elastic
micro-channels and micro-valves \cite{CFMtest, FPVAtest}. 

A simpler technology for designing CFMBs is to use free-flowing fluid through rigid channels, where no control layer or valves are needed~\cite{Banerjee2017,O'Neill,2004_LOC_Neils,SerialDilution}. Free-flowing biochips only require a switching mechanism to
inject or stop fluids at the inlets (or input ports). The fluid flow through the micro-channels is fully controlled by their
global hydraulic resistance and injection velocities at input ports. Traditionally, these chips were more popular among biochemists, as
they are inexpensive, easy to fabricate,  and 
suitable for high-throughput applications. However, free-flowing chips are mostly fully customized and
cannot be re-programmed easily for general usage. Since there is almost no control on fluid navigation, it is not straightforward to implement protocols such as
dilution or mixture preparation, which require high degree of programmability, flexibility, and
precision on concentration factors. Additionally, the layout design of a free-flowing chip should satisfy certain hydrodynamical  requirements in order to avoid turbulence through the channels.

The ubiquity of sample preparation (SP), which refers to the task of preparing dilution or mixture of fluids in a certain ratio, is felt in almost all biochemical protocols. Many algorithms for automated sample preparation are known both for DMFBs \cite{ACM2015,CAD2010_DMRW,CAD2014,REMIA,Date_sudip} as well as for CMFBs \cite{2008_LOC_Kim,Philip2016,Xmas,Rotary,Date_sudip}. Most of the on-chip microfluidic modules support the traditional (1:1) mix-split model where two equal-volume fluids are mixed and then split into two equal parts. Dilution (mixing) of fluids to achieve a desired target concentration factor ({\em CF}) (or a ratio of {\em CF}s) is implemented using a sequence of such operations. Note that {\em CF} is expressed as a fraction 0 $\leq$ {\em CF} $\leq$ 1. Clearly, the {\em CF} of the resulting mixture following one (1:1) mix operation is the arithmetic mean of the two parent-{\em CF}s. As a consequence, in most of the SP-algorithms, a target-{\em CF} is first approximated as an $n$-bit {\em binary} fraction, where $n$ is a positive integer that determines the desired accuracy of {\em CF}s.  The reliability of sample preparation depends on the correctness of (1:1) mixing steps and balancing of split operations during implementation~\cite{ACMSudip,error_oblivious, BanerjeeACM2020,BanerjeeVLSID2019,BanerjeeCBT}. SP-algorithms usually aim to minimize the number of mix-split operations (assay-time)~\cite{CAD2010_DMRW}, reactant-usage (cost)~\cite{REMIA}, or waste production~\cite{WARA}, and their complexities largely depend on the value of $n$, the required output volume, and the availability of physical resources on-chip. 

Free-flowing biochips are easy to fabricate and more reliable as only a few control ports are used. However, for these biochips, SP-algorithms are mostly unexplored because in the absence control valves, volume-metering during mixing or splitting of bulk fluid
is extremely difficult. No electronic design automation (EDA) tool is currently available that is capable of synthesizing such chips. One has to model several fluidic parameters (e.g., density, viscosity, inertial force, etc) in order to understand the underlying fluid dynamics. Moreover, the functionality of such chips is highly sensitive to the dimension of microfluidic channels, input pressure, and the overall fabric of the network, which determines the global hydraulic resistance, and in turn, the output flow.  All such limitations slow down the cycle of “Design-Prototyping-Testing”, thereby increasing the cost of such chips and turn-around-time. 

In this paper, our contribution is two-fold: (i) first, we develop a new SP-algorithm for producing dilution that can be easily emulated with a free-flowing chip, and (ii) present a methodology for designing the chip with details of its physical design. This chip is programmable in the sense that by choosing appropriate fluid inputs, we can produce an output flow with a desired concentration factor.  

The proposed algorithm introduces a non-traditional mixing model for the first time, where target-{\em CF}s are represented by rational numbers based on Farey sequence~\cite{Farey,StatisticsInFarey,FareyPolygons} instead of using classical binary fractions (i.e., where the denominator is a power of two only).  We show that such a model supports different fluid-mixing strategies, which are suitable for implementation with a free-flowing biochip. We next construct a 3D-model of the proposed channel architecture and study its behavior via COMSOL simulation framework~\cite{COMSOL}. The chip does not need any fluidic split operation and thus errors due to unbalanced splitting are completely avoided. Unused fluids are recycled efficiently and the output flow achieves quick convergence to the desired target-{\em CF}. The accuracy of {\em CF} obtained by this method compares favorably with prior art.

As stated above, we introduce a new method to discretize the {\em CF}-range using a Farey sequence $F_m$ of order $m$, which is an increasing sequence of all irreducible rational numbers $\frac{a}{b}$, where $a$, $b$, and $m$ are positive integers, and $0 < a < b\leq$ $m$~\cite{Farey,StatisticsInFarey,FareyPolygons}. The two boundary values $0$ and $1$ are also included in the sequence. 
Thus, given a value of a positive integer $m$, this allows us to approximate a target-{\em CF} using a fraction $\frac{a}{b}$. The value of $m$ can be chosen depending on the accuracy level in {\em CF} as close to that used in traditional methods such as \emph{twoWayMix} \cite{MinMix} or \emph{REMIA} \cite{REMIA}. It is known that the number of elements in $F_m$ asymptotically approaches to $\frac{3m^2}{{\pi}^2}$ \cite{Farey}, and thus the closed interval [0, 1] for the {\em CF}-range can be discretized by choosing a suitable subset of nearly equi-spaced fractions among the elements of the Farey sequence of a given order. Based on the above discretization scheme, we propose a generalized mixing model ($p : q$) and a new dilution algorithm namely {\em FSD} (Farey-Sequence based Dilution). In order to perform homogeneous mixing operations, we used flow-equalizers followed
by inter-digitized orthogonal obstacles inside the microchannels. Such
microfluidic channels can be easily manufactured using 
available techniques \cite{3Dserpentine}. We validate the
functionality of the network using COMSOL Multiphysics Software~\cite{COMSOL} and report performance results for several random and real-life test-cases.

The rest of the paper is organised as follows: Section \ref{DoP} presents the basics of sample preparation; Section~\ref{rworks} describes prior-work related to this research. We explain the
motivation behind this work in Section~\ref{motiv}. The proposed methodology is discussed in Section \ref{method} followed by architectural design detailing the fluidic network in Section \ref{AlgoLayout}. Simulation framework, experimental results on random and real-life target-{\em CF}s, and comparisons with earlier approaches are reported in Section \ref{ExpRes}. Finally, Section \ref{Concl} draws the conclusion.

%% file: Text/DoP.tex
\input{Text/GenMixing.tex}

%% file: Text/GenMixing.tex
In dilution, a raw sample fluid is mixed with a buffer fluid in a certain volumetric ratio to achieve desired target concentration factors (\emph{CF}, $0 \leq$ \emph{CF} $\leq 1$) of a sample, where the \emph{CF} of a raw sample (buffer) is considered as 1 (0) \cite{CAD2010_DMRW,MinMix}. In the ($p : q$) 
model, a mixing operation is 
performed between $p$-unit and $q$-unit volume fluids 
with \emph{CF} = $C_1$ and $C_2$, respectively. Thus, the \emph{CF} of the resultant mixture becomes $\frac{p \cdot C_1 + q \cdot C_2}{p + q}$. It reduces to the classical ($1 : 1$) mix-split model when the value of $p$ and $q$ become 1. Given a target-{\em CF}, a specific sequence
 of mix-split operations need to be performed (utilizing sample and buffer fluids) to achieve the desired target-{\em CF} of a sample. The complete sequence of such mix-split operations can be graphically visualized with a directed acyclic-graph called mixing tree/dilution graph~\cite{MinMix,CAD2010_DMRW}. The depth $d$ ($d > 0$) of the mixing tree represents the accuracy $n$  ($n > 0$) of a target-\emph{CF}, which is determined by a user-specified error-tolerance limit $\epsilon$ ($0 \leq \epsilon < 1$). In order to limit the \emph{CF}-error in the target-droplet by $\frac{1}{2^{n+1}}$,  each \emph{CF} is represented as $\frac{x}{2^n}$ ($n$ is a positive integer, and $x$ is a non-negative integer such that $0 \leq x\leq 2^n$), for the special case of the  $(p:q)$ mixing model where $p = q$. 

%% file: Text/related_work.tex
\noindent
SP-algorithms for implementation on valve-based CFMB chips were proposed in~\cite{Philip2016, Rotary} where valve-states (open/close) need to be controlled online following pre-scheduled dynamics. Albeit they offer the advantage of programmability to some extent, malfunctioning of control-valves may lead to errors in fluidic operations or performance degradation. For example,
valve expansion might affect the pressure gradient in the microchannels
of PDMS based CFMBs. Asynchronous valve actuations
may cause errors in fluidic operations and slow down the response
time~\cite{ValveMinimization, CtrlRoutingPinMinimization}. Any defect in elastic micro-valves or micro-pumps
may also introduce various microfluidic errors in valve-based CFMBs~\cite{CFMtest, FPVAtest}. Free-flowing programmable
architectures that obviate the need for valves as well as split operations were proposed in~\cite{Banerjee2017,BanerjeeCBT,BanerjeeVLSID2019} to overcome these limitations. The work in~\cite{Banerjee2017} studies a 2D microfluidic network for implementing dilution on a free-flowing biochip. In order to capture subtleties of mixing and diffusion phenomena
 in a more realistic fashion~\cite{2D3D}, a 3D-model for the fluidic network was presented later~\cite{BanerjeeVLSID2019}. These designs, however, need variable injection-velocities at input ports and precise timing sequence for fluid injection, which again are hard to implement in practice.  Free-flowing biochips resembling a binary tree was proposed in~\cite{BanerjeeCBT} to mitigate these problems. However, on the negative side, they require a special set of concentration values to be fed as inputs.


%% file: Text/motivation.tex
\noindent
The impact of several parameters that govern fluid flow through a network of channels needs to be considered while designing the channel structures and physical layout of a free-flowing biochip. Examples of such parameters include (i) density, (ii) dynamic and kinematic viscosities, (iii) fluid-injection velocities at input ports, and (iv) the volume of fluid injected into the system per unit time. Moreover,
in sample preparation (e.g., dilution), some of these parameters may also depend on the \emph{CF} of the solution. Hence, the hydraulic resistance, and consequently, the rate of fluid flow as well as miscibility also depend on the \emph{CF}-value of the fluid. 

Note that the volume and injection velocity of fluids in a channel network can be controlled by choosing appropriate physical dimensions of input ports and  input pressure. In previous approaches to preparing dilution with free-flowing biochips \cite{Banerjee2017,BanerjeeVLSID2019}, a large volume of fluids is wasted initially before the stablity of flow is reached at the sink-outlet(s). Also, fluid-injection velocities at different input ports along with their corresponding injection times, play a vital role in achieving the stability and accuracy of target-\emph{CF}. Adjustment of such velocities and timing is a challenging task \cite{Banerjee2017,BanerjeeVLSID2019}.

In digital microfluidics as well as in valve-based {CFMB}s, volumetric errors that often occur during split operations cause errors in target-\emph{CF}s \cite{ACMSudip,BanerjeeACM2020,error_oblivious}. Additionally, split operations may lead to the wastage of costly fluids. In free-flowing chips, this problem is avoided at the cost of consuming additional input fluids and producing a larger amount of target fluids. The combination of density, dynamic viscosity, fluid-velocity, and channel length determines the Reynolds number \cite{Re_NS} of the flow, which in turn, dictates the nature of fluid-motion (such as turbulance or laminar). 

Two issues thus need to be addressed while designing free-flowing {CFMB}s:
\begin{itemize}
\renewcommand{\labelitemi}{\scriptsize$\blacksquare$}
\item off-line issues such as the selection of underlying dilution algorithms, physical layout of the chip, optimization parameters, design complexities; 

\item online issues such as handling of microvalves, the choice of injection velocities and timing management so as to achieve faster stability of flow and desired accuracy of target-\emph{CF}.
\end{itemize}

In the proposed approach, we need to set the
states of each inlet port {\em a-priori} (offline), i.e., before
executing the dilution assay on the free-flowing biochip. Thus, it does not require any real-time control of ports. We also  present a 
new architecture based on Farey sequence that supports programmability, and at the same time, obviates the need for different injection velocities and precise online injection timing. This improves upon other free-flowing fluidic architectures proposed earlier~\cite{Banerjee2017,BanerjeeCBT,BanerjeeVLSID2019}.  We construct a 3D-model of the proposed channel architecture considering shear stress between the wall and fluid layer as well as those between adjacent fluid layers in both parallel and perpendicular directions, and study its behavior via {COMSOL} simulation framework. 


%% file: Text/Proposed_method.tex
\noindent

In this section, we will describe a new methodology for designing the
free-flowing 3$D$ microfludic biochips for sample preparation.
\vspace{-7pt}
\subsection{Representation of CF-values using Farey sequence}

\label{lab:gen_fsd}

\begin{figure*}[t]
	\centering
	\frame{\includegraphics[width=16cm,height=4.3cm]{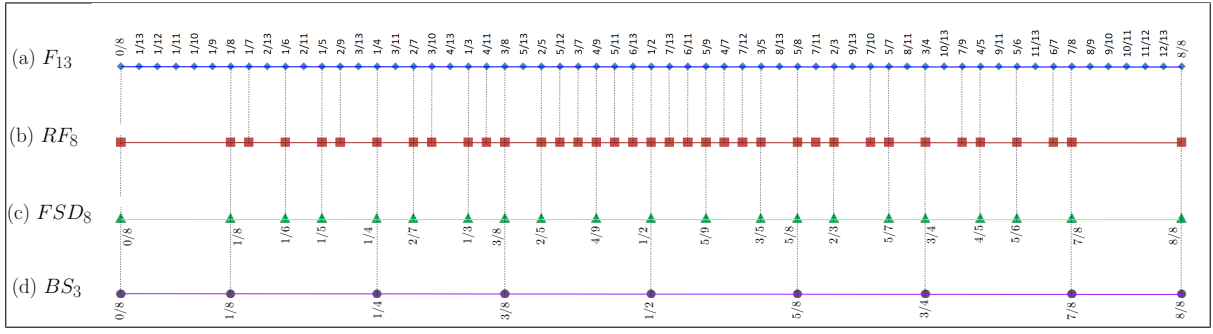}}
	\caption{Derivation of $FSD_{8}$ sequence from Farey sequence ($F_{13}$).}
	\label{fig:FSD4mFarey}
\end{figure*}

\begin{figure}[!ht]
	\centering
	\includegraphics[width=8cm,keepaspectratio]{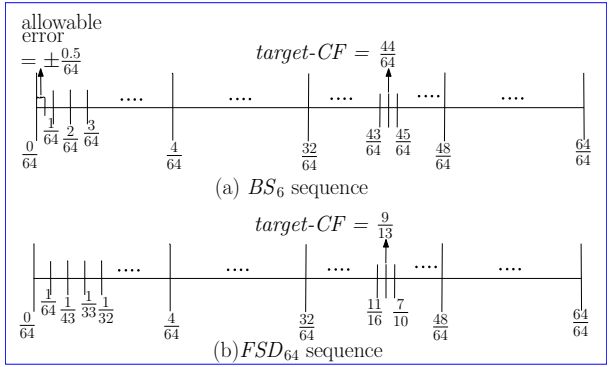}

	\caption{(a) Representable $CF$-values for $n = 6$ used in traditional SP-Algorithms;
 (b) by the proposed method, which offers more dense discretization for the same value of $n$.}
	\label{fig:halfError}
\end{figure}

As stated earlier, in almost all prior state-of-the-art dilution algorithms both in {DMFB} and {CFMB} domains, each target-\emph{CF} (\emph{tCF}) is represented as $\frac{x}{2^n}$, where 2 appears in the denominator because of the underlying $(p : p)$ mixing-model; $n$ is a positive integer. Thus, the \emph{CF}-values ranging from 0\% to 100\% are discretized as $\frac{0}{2^n}$, $\frac{1}{2^n}$, $\cdots$ $\frac{2^{n-1}}{2^n}$, $\frac{2^n}{2^n}$,  $n$ denoting the accuracy level~\cite{Banerjee2017,BanerjeeVLSID2019,BanerjeeCBT}. We call this as binarized sequence \textit{BS} of order $n$ (i.e., \textit{$BS_n$}). For example, the possible target-\emph{CF} values of a binarized sequence of order 3 (\textit{$BS_3$}) are shown in the bottom row in Fig. \ref{fig:FSD4mFarey}(a) for accuracy level = $3$. It can be seen that the target-\emph{CF}s appear following a gap of $\frac{1}{2^n}$ implying that  the \emph{CF}-error in the target-\emph{CF} lies within $\pm \frac{0.5}{2^n}$. Therefore, the target-\emph{CF} = $\frac{ 44.375}{64}$ (see Fig. \ref{fig:halfError}(b)) will be approximated as $\frac{44}{64}$. Hence, the error in the target-\emph{CF} becomes $\frac{-0.375}{64}$.


In this work, we relax the constraint on the denominator part of each \emph{tCF}, and propose a dilution algorithm for flow-based biochips based on Farey-sequence arithmetic\cite{Farey,FareyPB}. More precisely, we represent each target-\emph{CF} using an element of a Farey-sequence. The Farey sequence~\cite{Farey} of order $m$ = $1$ to $8$ are listed below:

{\noindent
$F_1$ = \{$\frac{0}{1}, \frac{1}{1}$\},\\
$F_2$ = \{$\frac{0}{1}, \frac{1}{2}, \frac{1}{1}$\},\\
$F_3$ = \{$\frac{0}{1}, \frac{1}{3}, \frac{1}{2}, \frac{2}{3},$ $\frac{1}{1}$ \},\\
$F_4$ = \{$\frac{0}{1}, \frac{1}{4}, \frac{1}{3}, \frac{1}{2}, \frac{2}{3}, \frac{3}{4}, \frac{1}{1}$ \}, \\
$F_5$ = \{$\frac{0}{1}, \frac{1}{5}, \frac{1}{4}, \frac{1}{3}, \frac{2}{5}, \frac{1}{2}, \frac{3}{5}, \frac{2}{3}, \frac{3}{4}, \frac{4}{5}, \frac{1}{1}$ \}, \\
$F_6$ = \{$\frac{0}{1}, \frac{1}{6}, \frac{1}{5}, \frac{1}{4}, \frac{1}{3}, \frac{2}{5}, \frac{1}{2}, \frac{3}{5}, \frac{2}{3},$ $\frac{3}{4},$ $\frac{4}{5}, \frac{5}{6}, \frac{1}{1}$ \},\\
$F_7$ = \{$\frac{0}{1}, \frac{1}{7}, \frac{1}{6}, \frac{1}{5}, \frac{1}{4}, \frac{2}{7}, \frac{1}{3}, \frac{2}{5}, \frac{3}{7}, \frac{1}{2}, \frac{4}{7}, \frac{3}{5}, \frac{2}{3}, \frac{5}{7}, \frac{3}{4}, \frac{4}{5}, \frac{5}{6}, \frac{6}{7}, \frac{1}{1}$ \}, \\
\mbox{$F_8$ = \{$\frac{0}{1}, \frac{1}{8}, \frac{1}{7}, \frac{1}{6}, \frac{1}{5}, \frac{1}{4}, \frac{2}{7}, \frac{1}{3},$ $\frac{3}{8}, \frac{2}{5}, \frac{3}{7}, \frac{1}{2}, \frac{4}{7}, \frac{3}{5}, \frac{5}{8}, \frac{2}{3}, \frac{5}{7}, \frac{3}{4}, \frac{4}{5}, \frac{5}{6}, \frac{6}{7}, \frac{7}{8}, \frac{1}{1}$\}.}
}

For example, the set of \emph{CF}s derived from $F_{13}$ is shown in Fig. \ref{fig:FSD4mFarey}(a). It has been observed from Fig. \ref{fig:FSD4mFarey}(c) that the number of representable \emph{CF}-values by the proposed method is much higher than binarized discretization for accuracy level = $3$. Thus, for target-\emph{CF} = $\frac{44.375}{64}$, the approximated \emph{CF} becomes $\frac{9}{13}$ (Fig.~\ref{fig:halfError} (b)), and hence, the error in \emph{CF} becomes smaller (i.e., $\frac{0.067}{64}$) compared to binarized discretization.

The rationale behind choosing a value for $m$ lies in the physical design of the proposed diluter architecture (shown in Fig.~\ref{fig:3D_layout}) which will have $n$ inlets for injecting the sample fluid and $n$ inlets for injecting the buffer fluid. Furthermore, the cross-section of these inlets will be progressively doubled. Thus, the width of the rectangular inlets would be 
$1x$, $2x$, $2^2x$, $2^3x$, $...$, $2^{n-1}x$ ($x$ represents the unit). Hence, both for sample and buffer fluids, the minimum (maximum) amount that can be injected into the channel is $0$ ($2^n$ - $1$) units.
Now, starting from the \emph{Farey sequence}  $F_{2^{n+1} - 3}$, we can construct a shorter sequence yet having a larger number of terms than those in \emph{BS}$_n$. 
 We name this as \emph{FSD}$_{2^n}$. Note that \emph{FSD}$_{2^n}$ is a proper subset of $F_m$, where $m$ = $2^{n +1} -3$. Algorithm \ref{algo:FSDGen} described later how to construct \emph{FSD}$_{2^n}$ from \emph{F}$_{2^{n +1} -3}$.
\begin{algorithm}[!ht]
	\footnotesize
	\DontPrintSemicolon 
	\let\oldnl\nl
	\newcommand{\nonl}{\renewcommand{\nl}{\let\nl\oldnl}}%
	\KwIn{Order (accuracy) \textit{n} ($\geq$ 1) of $BS_{n}$}
	\KwOut{{\em FSD} sequence of order $2^n$.
	}
	Generate the general sequence of {\em CF}s ($BS_n$) of order $n$ ; i.e.; $BS_{n}$ = \{$\frac{0}{2^{n}}$, $\frac{1}{2^{n}}$, $\frac{2}{2^{n}}$, .... , $\frac{2^{n-1}}{2^{n}}$, $\frac{2^{n}}{2^{n}}$\}\;
	
	Generate the numbers following Farey sequence based arithmetic of order $m$ ($F_m$), where $m$ = $2^{n+1}-3$ ; i.e.; $F_{m}$ = \{$\frac{0}{m}$, $\frac{1}{m-1}$, $\frac{1}{m-2}$, .... , $\frac{m-1}{m}$, $\frac{m}{m}$\}\;
	
	Replace boundary values of $F_m$, i.e., \{$\frac{0}{1}$\} and \{$\frac{1}{1}$\}
	with the values of buffer ($\frac{0}{2^n}$) and sample ($\frac{2^n}{2^n}$), respectively;

	Select a set of numbers $RF$ (of order ${2^{n}}$) from $F_m$ (i.e., $RF_{2^{n}}$ $\subset$ $F_m$) as follows:\;
	\nonl \hspace{0.5mm}(a) $RF_{2^{n}}$ = \{$x \mid x \in$ \{$\frac{0}{2^{n}}$, $\frac{\phi}{\psi}$,  $\frac{2^{n}}{2^{n}}$\} \},
	 where $\frac{\phi}{\psi}$ $\in$ $F_m$, (0 $ < (\phi$, $\psi$) $<$ $2^{n}$), and (($\psi$ - $\phi$) $<$ $2^{n}$)\;
	
	Select a set of numbers $FSD_{2^n}$ from $RF_{2^{n}}$ (i.e., $FSD_{2^{n}}$ $\subset$ $RF_{2^{n}}$) based on: \;
	\nonl \hspace{0.5mm}(a) $FSD_{2^{n}}$ = \{$x \mid x \in$ $\frac{0}{2^{n}}$, $\frac{\phi}{\psi}$, $\frac{2^{n}}{2^{n}}$\},
	where $\frac{\phi}{\psi}$ $\in$ \{$BS_n$\} $\bigcup$ \{$\frac{y}{2^{n}}$ $<$ $\frac{\phi}{\psi}$ $<$ $\frac{2y+1}{2^{n+1}}$$\mid$ where $\phi$ is minimum among all numerators of ($\frac{y}{2^{n}}$,$\frac{2y+1}{2^{n+1}}$)\}  $\bigcup$ \{$\frac{2y+1}{2^{n+1}}$ $<$ $\frac{\phi}{\psi}$ $<$ $\frac{y+1}{2^{n}}$$\mid$ where $\phi$ is minimum among all numerators of ($\frac{2y+1}{2^{n+1}}$,$\frac{y+1}{2^{n}}$)\} 
	
	\Return{ $FSD_{2^{n}}$}\;
	
	\caption{Generation of $FSD_{2^n}$.}
	\label{algo:FSDGen}
\end{algorithm}
In the beginning (Line 1), Algorithm~\ref{algo:FSDGen} creates the binarized sequence
of {\em CF}-values \textit{$BS_n$} (Fig.~\ref{fig:FSD4mFarey} (d)). In Line 2, all elements of 
Farey sequence of order $m$ ($F_m$) are created.
In Line 3,
the boundary values of $F_m$,
i.e., \{$\frac{0}{1}$\} and \{$\frac{1}{1}$\} are replaced with $\frac{0}{2^n}$ (buffer {\em CF}) and $\frac{2^n}{2^n}$ (sample {\em CF}), respectively. For $n = 3$, we have shown a Farey sequence of order 13 ($F_{13} = 2^{(3+1)} - 3$)
in Fig.~\ref{fig:FSD4mFarey} (a) which comprises 59 elements. First, the fractions are eliminated for which the numerators are greater than 7 (=$2^3$ - 1) or the difference
between the numerator and the denominator exceeds 13 (=$2^{(3+1)}$ - 3). This set is called 
reduced Farey sequence ($RF_8$) as shown in Fig.~\ref{fig:FSD4mFarey} (b)
(Step 3). This reduced Farey sequence is equivalent to \{$\frac{a}{a+b}$ $|$ $a,$ $b$ $\in$ \{$0, 1, \cdots, 2^{n-1}\}$\}, where $a$ and $b$ are positive integers such that $0 \le a$, $b \le 2^n - 1$, ($a + b$) $\not= 0$. Note that the underlying architecture exactly supports  the same sequence: that is mixing of $a$ units of sample with $b$ units of buffer producing target-\emph{CF} = $\frac{a}{a+b}$. After execution of Step 4, we form the sequence  $FSD_8$. Note that $BS_3$  has fewer elements than $FSD_8$ between the open interval (0, 1).
In general, first include all numbers of \textit{$BS_n$} into $FSD_{2^n}$. Then, select a set of numbers from
$RF_{2^n}$ within the range of ($\frac{1}{2^n}$,  
$\frac{2^{n-1}}{2^n}$) and insert them into $FSD_{2^n}$ as follows. Suppose
two consecutive elements of $FSD_{2^n}$ are ($\frac{y}{2^n}$,  
$\frac{y+1}{2^n}$) ($y = 1,2,3,\cdots 2^{n-2}$). Then, the algorithm selects
two numbers 
(among a set of numbers between every two consecutive numbers ($\frac{y}{2^n}$, $\frac{y+1}{2^n}$)) from $RF_{2^n}$ and insert them into $FSD_{2^n}$. Note that one number, with minimum numerator value
is selected within the range ($\frac{y}{2^n}$, $\frac{2y+1}{2^{n+1}}$) and
another with minimum numerator value is selected within the range
($\frac{2y+1}{2^{n+1}}$, $\frac{y+1}{2^{n+1}}$). Here, minimum values are
selected for reducing the sample consumption. All {\em CF}-values of $FSD_{8}$ generated by the proposed method is shown in Fig.~\ref{fig:FSD4mFarey} (c).
Thus, for the same {\em CF}-range, the proposed method generates more {\em CF}-values (21) compared to binary methods (8 {\em CF}-values for $n = 3$). Hence, the proposed method approximates a given {\em CF} with higher accuracy for a given value of $n$. Here, another interesting thing is that $RF_{2^n}$ = \{$\frac{a}{a+b}$ $|$ $a,$ $b$ $\in$ \{$0, 1, \cdots, 2^{n-1}$\} is a sub-sequence of Farey sequence \emph{F}$_{2^{n +1} -3}$ and \emph{FSD}$_{2^n}$ is a sub-sequence of $RF_{2^n}$. 

\begin{theorem}
	The numerator of each irreducible fraction element in ${FSD}_{2^n}$ is less than $2^n$. Also, the difference between denominator and numerator is also less than $2^n$.
\end{theorem}
~~~\textbf{Proof:}
    Numerator of each $\frac{\phi}{\psi}$ (i.e., $\phi$) must be within the open interval (0, $2^n$). 
	In the proposed diluter network, the activation of all sample inlets at a time can inject 
	$\sum_{i = 0}^{n-1}2^i$ = ($2^n-1$) units of sample fluid. Thus 
	$\phi$ $<$ $2^n$. 
	Also, $\phi$ $>$ $0$, since some units of sample fluid should be present in the dilution. The difference between denominator $ \psi$ and numerator $\phi$ represents the number of buffer fluid units used in the dilution to achieve \emph{tCF} = $\frac{\phi}{\psi}$. 
	Note that the activation of all buffer inlets at a time can inject 
	$\sum_{i = 0}^{n-1}2^i$ = ($2^n-1$) buffer units. Thus, ($\psi$ - $\phi$) $<$ $2^n$. 

\begin{theorem}
   $|{FSD}_{2^n}| > |{BS}_n|$, i.e., the number of elements in ${FSD}_{2^n}$ is always greater than the number of elements in ${BS}_n$.
\end{theorem}

\textbf{Proof:} This follows from step $5$ of Algorithm \ref{algo:FSDGen}. Let us consider two consecutive elements from $BS_n$ sequence $\frac{t}{2^n}$ and $\frac{t+1}{2^n}$. The middle element of these two is $\frac{2t+1}{2^{n+1}}$ (= \{$\frac{t}{2^n}$ + $\frac{t+1}{2^n}$\}/2). Next, we choose $\frac{\phi}{\psi}$ in such a way that the numerator $\phi$ is minimum among all those lying between the open interval of the lower half, i.e., ($\frac{t}{2^{n}}$,$\frac{2t+1}{2^{n+1}}$). 
Find another $\frac{\phi}{\psi}$ where the numerator $\phi$ is minimum among all those lying between the  open interval of the upper half, i.e., ($\frac{2t+1}{2^{n+1}}$,$\frac{t+1}{2^{n}}$). Thus $|BS_n|$ = $2^n$ and $|FSD_{2^n}|$ = $3 (2^{n} - 1)$.
	 

\begin{theorem}
The set ${FSD}_{2^n}$ includes the set ${BS}_n$, i.e., \emph{FSD}$_{2^n}$ $\supset$ \emph{BS}$_n$.
\end{theorem}

\textbf{Proof:}
Follows easily from the construction of $FSD_{2^n}$ as described in Algorithm~\ref{algo:FSDGen}.
\subsection{Mixing tree based on Farey-approximation of {tCF}}
\label{lab:gen_mix_tree}

The proposed method first maps the target-{\em CF} (in the form $\frac{\phi}{\psi}$) to its nearest element of farey sequence $FSD_{2^n}$ for creating the mixing tree (which represents the sequence of mixing steps for achieving the desired target-{\em CF}). In order to accomplish this task, it adopts the \textit{FindClosestFast} technique~\cite{FareyPB}.  For example, if we invoke {\it FindClosestFast}, the nearest element of  \emph{tCF} = $\frac{44.375}{64}$ (= 69.3\%) becomes $\frac{9}{13}$ (i.e., with {\em CF}-error = $\frac{0.067}{64}$) for  $n = 6$, as shown in Fig.~\ref{fig:halfError} (b). Here, we skip the algorithm details due to the limitation of allowable pages.

\begin{figure}[h]
	\centering
	\includegraphics[width=8cm,height=3.8cm]{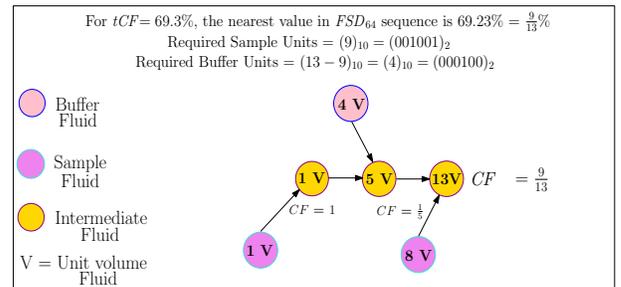}
	
	\caption{{\em FSD} tree for \emph{tCF}$ = $$\frac{9}{13}$. }
	\label{fig1:FSD tree}
\end{figure}

\begin{figure}[h]
    \centering
    \fbox{\includegraphics[width=8cm,keepaspectratio]{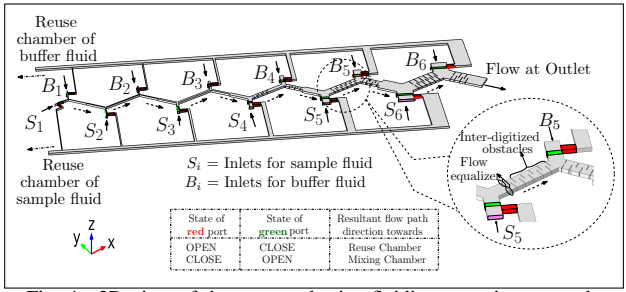}}
    \caption{3D-view of the proposed microfluidic serpentine network.}
    \label{fig:3D_layout}
\end{figure}

We will now demonstrate how given a target-{\em CF}, Farey arithmetic
can be used to determine the required sequence of mixing operations. 
Fig. \ref{fig1:FSD tree} shows the mixing tree to achieve the target-{\em CF}$= \frac{9}{13}$.
In order to produce this target concentration, we need to mix 9 units of sample/reactant with 4 (= 13-9) units of buffer. We design a fluidic network that comprises six inlets each for injecting sample and buffer as shown in Fig.~\ref{fig:3D_layout}.
Also, the cross-sections of the inlet pipes are progressively doubled. The inlets can be made open or closed under external control as desired.
For this target-{\em CF}, 6-bit binary representations of $\phi$ $(9)_{10}$ and $\varphi$ $(4)_{10}$ become $(001001)_2$ and $(000100)_2$, respectively. The states (fluid-in/closed) of the inlet ports from left-to-right in Fig.~\ref{fig:3D_layout} needed to feed (stop) sample (buffer) into the mixing channel correspond to the respecting bits of $\phi$ ($\varphi$) as seen from the right-to-left. The progressive doubling of cross-sections of inlet ports from left-to-right takes care of the positional weights of the binary bits. In other words, by controlling the inlet ports, one can enable $\phi$ ($\varphi$) units of sample (buffer) to enter into the mixing channel so that the desired target concentration is produced at the output port.

%% file: Text/AlgoLayout.tex
\input{Text/ProposedApproach.tex}

%% file: Text/ProposedApproach.tex

\subsection{Proposed Layout}
\begin{figure}[!h]
  \centering
  
    \begin{tabular}{cc}
    \multicolumn{2}{c}{\frame{\includegraphics[width=8.4cm,height = 4.0cm]{pics/8_BDP_zoneMarked.pdf}}} \\
    \multicolumn{2}{c}{(a)} \\
        \frame{\includegraphics[width=4cm,keepaspectratio]{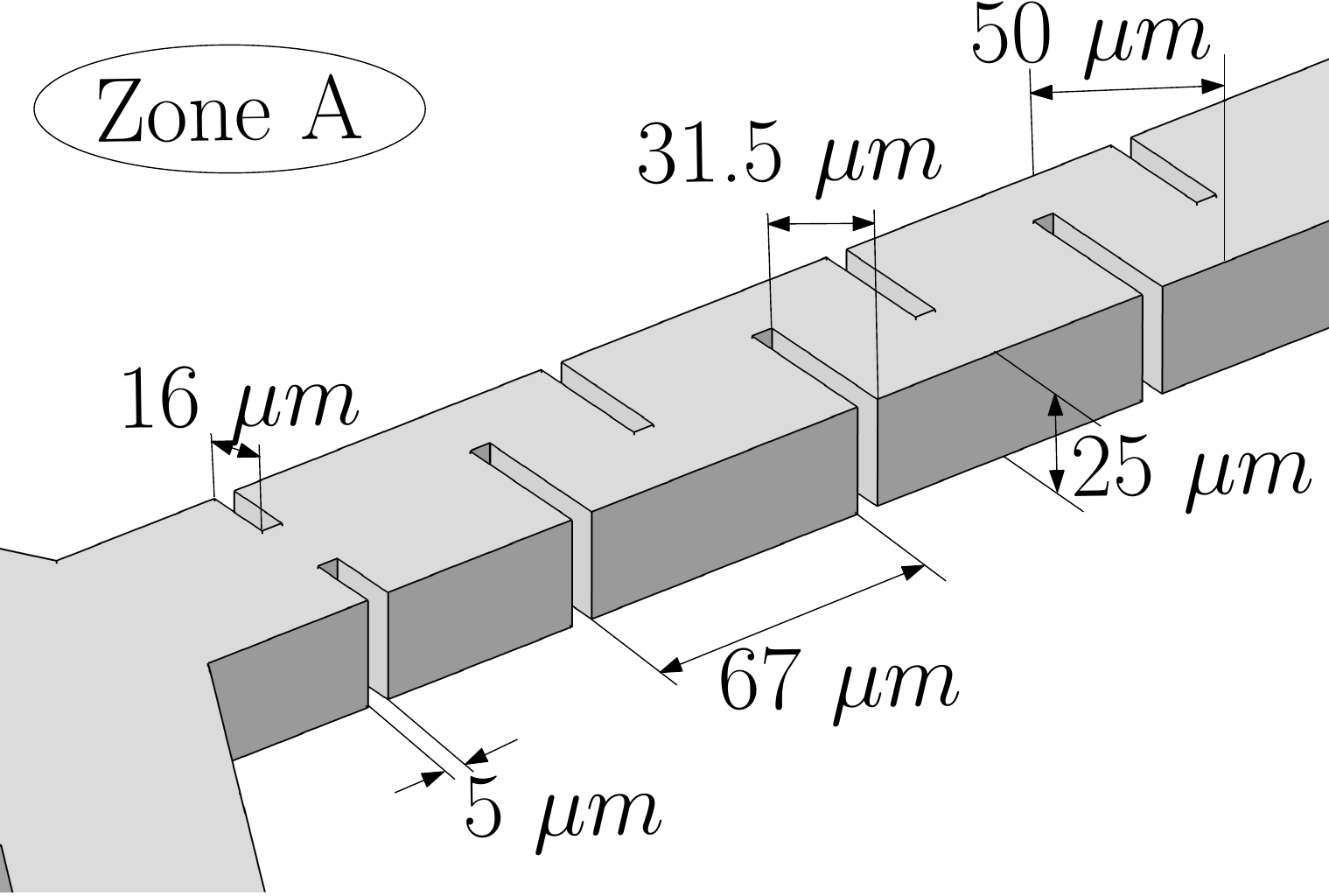}}  &  \frame{\includegraphics[width=4cm,keepaspectratio]{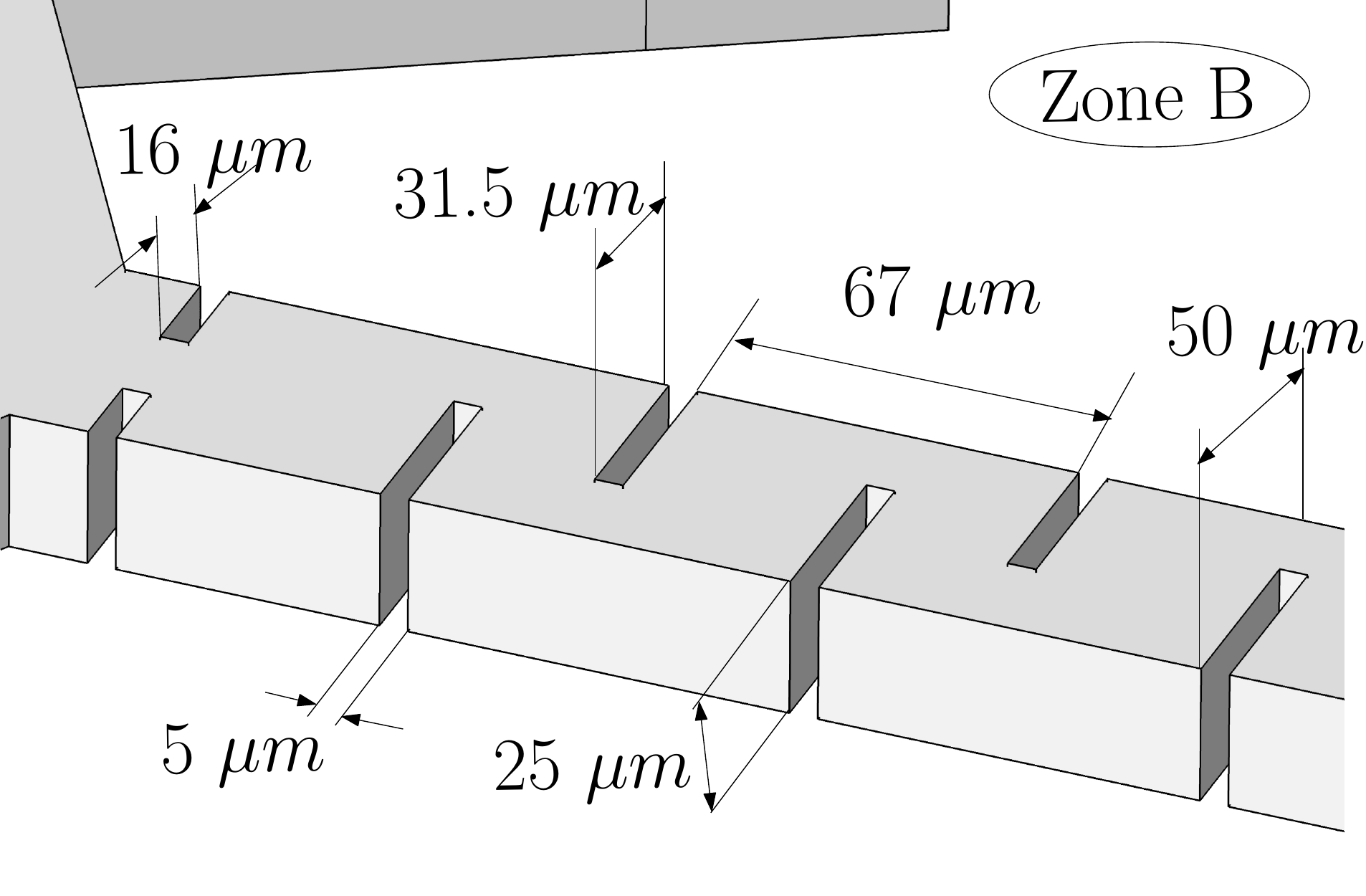}}\\
          (b) & (c) \\
        \frame{\includegraphics[width=4cm,keepaspectratio]{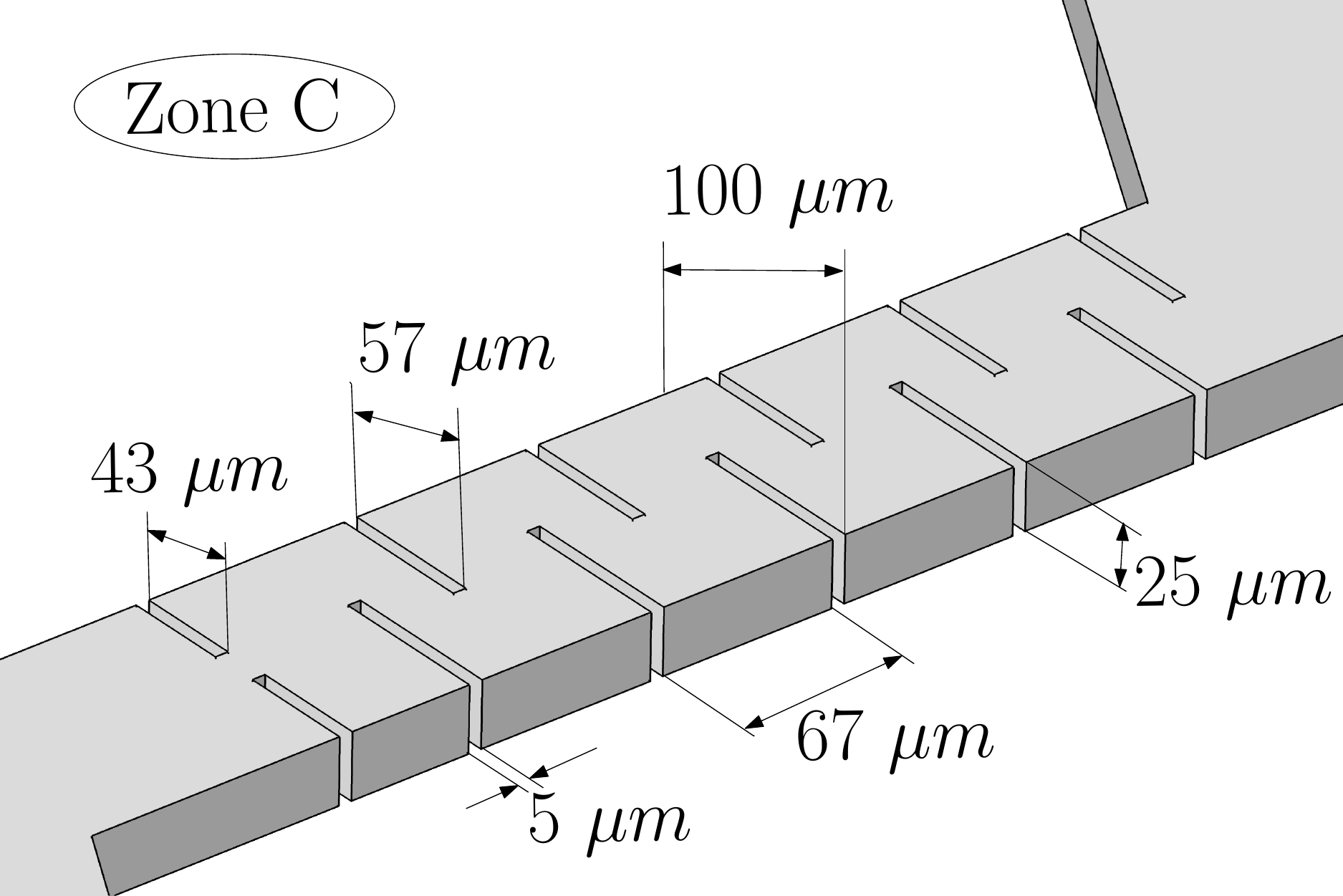}}  & \frame{\includegraphics[width=4cm,keepaspectratio]{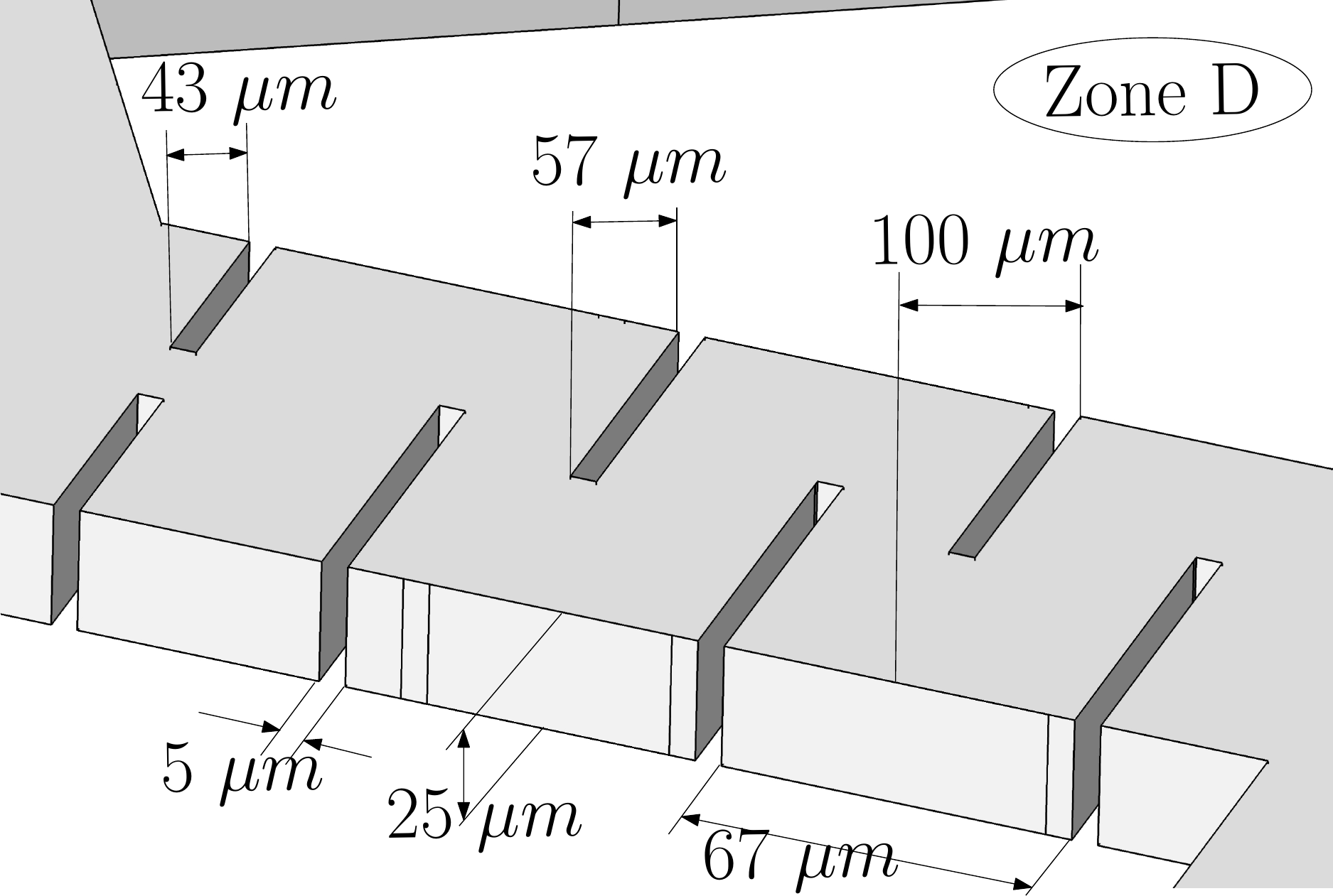}} \\
        (d)  &  (e)\\
         \frame{\includegraphics[width=4cm,keepaspectratio]{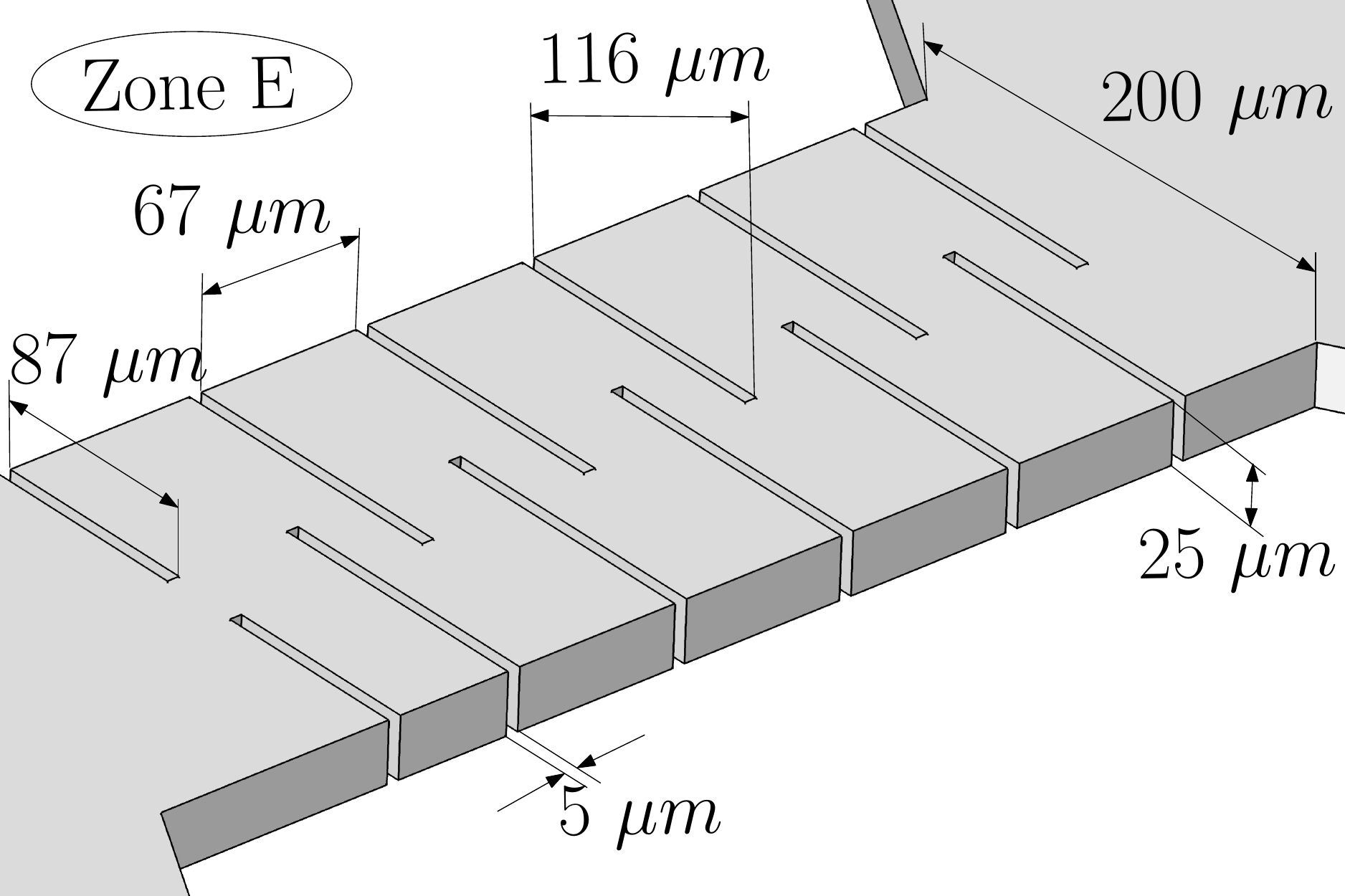}} & \frame{\includegraphics[width=4cm,keepaspectratio]{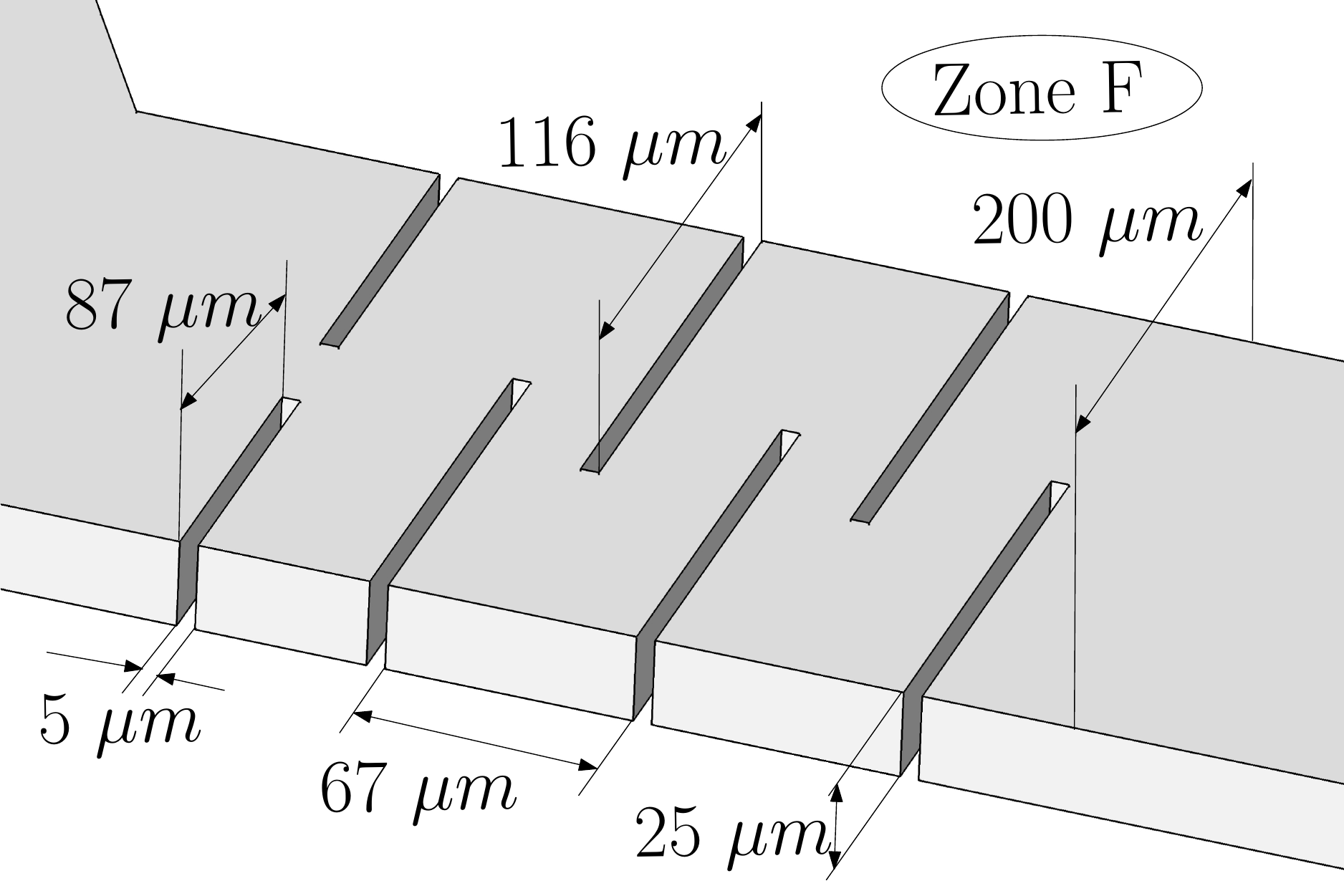}} \\
        (f)  &  (g)\\
    \end{tabular}%
     \caption{(a) Varying turning radii in the fluidic network;  (b-g) the geometry of inter-digitized obstacles in different zones of the channel network.}
  \label{fig:TurningRadius}%
\end{figure}

We will now describe the architecture and the layout design of the proposed free-flowing fluidic network, and required input assignments to produce a given target-$CF$. A three-dimensional model of the proposed fluidic network is presented that is tailored to implement the intrinsic subtleties of \emph{FSD}-algorithm. The design parameters of the network include height, width, and length of channels, the number of input and output ports and their cross-sections, layout geometry (how the channels are interconnected and resistive fins are shaped), and pressure-specs (fluid-injection velocities and timing). Most of the design parameters are pre-determined and invariant regardless of the target-{\em CF}. However, the number of input ports will depend on the desired accuracy-level ($n$) of target concentration. For demonstration purpose, we choose $n = 6$, thus the error in target-\emph{CF} will be less than $\frac{1}{128}$. The overall geometric layout of the corresponding architecture is shown in Fig. \ref{fig:3D_layout}, which has six inlet ports for injecting sample and buffer, each, and one output port, Thus, altogether it is a  $12$-input, 1-output free-flowing microfluidic network with no control valves in the mixing channel. The serpentine channel is built using inter-digitized obstacles that facilitate homogeneous mixing of fluids. The physical dimensions of the channel and obstacles are shown in Fig.~\ref{fig:TurningRadius} (b) - Fig.~\ref{fig:TurningRadius} (g).   

In order to perform homogeneous mixing operation, the layout uses  passive in-channel serpentine mixers equipped with inter-digitized fins~\cite{Review_Lee_SerpentineMixer1,SerpentineMixer3} with different turning radii as shown in Fig. \ref{fig:TurningRadius}.

The cross-sections of input ports are made progressively doubled so that the volume of fluid injected per unit time through them under constant pressure, increases two-fold at every consecutive inlets.
Also, a controlling port is placed at every inlet which may be kept open or closed depending on which input needs activation. Unlike other valve-based chips, these ports require only off-line control, i.e., they are set at the beginning of the assay depending on the target-\emph{CF}, and do not need any further control during the execution of the assay.
We keep the height of the entire channel network fixed at $25\mu m$. 
On application of constant pressure to all inlets, the rate of fluid flow (speed) will remain constant throughout the entire length of the mixing channel. 
The fluid-flux (volume flowing per unit time) increases in geometric progression along its path until the flow finally reaches the output port. We select input pressure in such a way that laminar flow involving miscible fluids is guaranteed through the channel network. 
Proposed method maintains laminar flow in the fluidic network by controlling
value of the Reynolds number ($\mathbb{R}$)~\cite{Re_NS} using
Equation~\ref{eq:R}:
\begin{equation}
\label{eq:R}
Reynolds \hspace{1mm} number \hspace{1mm} \mathbb{R} = \frac{Inertial \hspace{1mm} force}{Viscous \hspace{1mm} force} = \frac{\rho u l}{\mu}
\end{equation} 
A high (low) value of $\mathbb{R}$ indicates the existence of turbulent 
(laminar) flow in the network. In the proposed model, $\mathbb{R}$ is kept lower than 2000 by proper choice of $l$ (length of the channel), $u$ (fluid-injection rate) which assure 
laminar flow throughout the fluidic network for given values of $\rho$ (density of fluid) and $\mu$ (dynamic viscosity). 

\begin{figure}[t]
    \centering
    \fbox{\includegraphics[width=8.5cm,keepaspectratio]{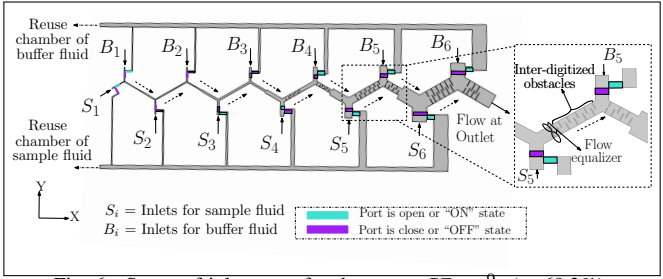}}
    \caption{States of inlet ports for the target \emph{CF} = $\frac{9}{13}$ ($\equiv$ 69.3\%).}
    \label{fig:Architecture}
\end{figure}

The proposed network architecture comprises two sets of flow paths. One of them corresponds to the regular mixing channel as described above. The other one (re-use channel) enables the sample and buffer fluids that are not needed at some inlets to return to their respective source reservoirs. As shown in Fig. \ref{fig:3D_layout} and Fig.~\ref{fig:Architecture}, an additional open/close port controller is placed on a by-pass channel attached to each inlet. Fluid injected at an inlet arm can flow through either the mixing channel or the re-use channel depending on the state of the corresponding port controller,
and their states are determined by the given target-\emph{CF}. 

Note that two different re-use channels are incorporated in the fluidic network – one for buffer-fluid outflow and the other one for sample-fluid outflow with a separate port attached to each inlet.  Ports attached towards the mixing channel  and those towards the re-use channel are operated in complementary fashion for each inlet (see Fig.~\ref{fig:Architecture}). The rationale behind  it is two-fold: first, they help equalize the hydraulic resistance confronted per unit area for every inlet; secondly, fluid-injection time through different inlets can be made concurrent irrespective of the target-\emph{CF}. 
The components of input fluids that are required in the dilution process are allowed to enter into the mixing chamber, whereas those not needed are fully diverted towards the re-use channel so that they can return to their respective sources. 

In order to enable balanced entry of fluids towards the mixing zone and to facilitate homogeneous diffusion-based mixing, flow-equalizers (septums) are used followed by a series of inter-digitized orthogonal obstacles (fins) in the mixing channel as shown in Fig. \ref{fig:TurningRadius}. The first pair of inlets used for injecting sample and buffer are oriented akin to a $Y$-junction with its two arms converging with a separation angle of $120^{\circ}$ ; it is then cascaded with a bisecting straight mixing channel. After that, each inlet (either carrying sample or buffer) arm is
joined with each of its incoming and outgoing mixing channels, with $120^{\circ}$  orientation (Fig. \ref{fig:3D_layout}).

\subsection{Procedure for generating a target {CF} using the proposed layout}

The network topology shown in Fig.~\ref{fig:Architecture} demonstrates the running
example target {\em CF} = $\frac{9}{13}$ ($\equiv$ 69.3\%). For each inlet port, the flow-controlling state should be set properly in order to produce a given
target-\emph{CF}. As discussed,  to achieve a target \emph{CF} $\frac{\phi}{\psi}$, $\phi$
unit of raw sample/reactant needs to be mixed with $\varphi(=\psi - \phi)$ unit
of buffer fluid. For example, for the target \emph{CF} $\frac{9}{13}$, the
6-bit binary forms of $\phi$ and $\varphi$ becomes 001001, and 000100, respectively. The weighted values of
these binary bits correspond to the amount of buffer and
sample to be injected, respectively.
Therefore, a bit 
value $1$ $(0)$ indicates that the corresponding inlet port towards the mixing chamber
is set as open/ON (closed/OFF) to let the fluid enter into the mixing chamber
(re-use branch). The controller attached to the re-use branch is set in opposite fashion. In Fig. \ref{fig:3D_layout}, an open (closed) port is colored as green (red). For the example target \emph{CF} = $\frac{9}{13}$, the states of the inlets towards the mixing chamber would be the following when viewed from left-to-right in Fig. \ref{fig:Architecture}:   sample: ({\em on, off, off, on, off, off}); buffer: ({\em off, off, on, off, off, off}). As stated earlier, the controllers towards the corresponding re-use branch are set in  opposite states. Depending on the target \emph{CF},  each inlet is set at the beginning of the dilution assay, 
and no further real-time control is required. 

\subsection{Throughput-sensitive choice of input flows}

The proposed device also has another useful feature: For a given target-\emph{CF}, the values of $a$ and $b$ can be chosen in different ways depending on the required output flow-rate. The minimum throughput corresponds to the elements of the \emph{FSD}-sequence, where each of them is an irreducible fraction. However, for other choices of $a$ and $b$ that yield the same value of $\frac{a}{a+b}$, we can obtain larger throughput.  Some examples are shown in Table \ref{tab:troughputSensitive}.
\setlength{\tabcolsep}{2.2pt}
\renewcommand{\arraystretch}{1.2}
\begin{table}[!h]
    \centering
    \caption{Throughput-sensitive Flow-Rate}
    \begin{tabular}{||c|c |c|c||}
    \hline 
         Target-\emph{CF}	&	Sample	&	Buffer	&	Rate of output-flow 	\\	\hline \hline
	&	1	&	2	&	3	\\	\cline{2-4}
$\frac{1}{3}$	&	2	&	4	&	6	\\	\cline{2-4}
	&	3	&	6	&	9	\\	
\hline
	&	1	&	3	&	4	\\	\cline{2-4}
$\frac{1}{4}$	&	2	&	6	&	8	\\	\cline{2-4}
	&	3	&	9	&	12	\\	
\hline
	&	4	&	1	&	5	\\	\cline{2-4}
	&	8	&	2	&	10	\\	\cline{2-4}
$\frac{4}{5}$	&	12	&	3	&	15	\\	\cline{2-4}
	&	16	&	4	&	20	\\	\cline{2-4}
	&   20  &   5   &   25  \\ \cline{2-4}
\hline
\end{tabular}
    
    \label{tab:troughputSensitive}
\end{table}

%% file: Text/ExpRes.tex
We have simulated the proposed fluidic network using COMSOL Multiphysics software \cite{COMSOL} and studied the performance of the \emph{FSD}-algorithm for various synthetic and real-life test-cases. We observe less than $1\%$ error in the target-\emph{CF} with respect to the theoretical \emph{CF} value in each instance.

\subsection{Simulation framework}
\input{Text/new_sim_framework}
\subsection{Results and Discussions}
\input{Text/Results.tex}

%% file: Text/new_sim_framework.tex
\noindent
During simulation, we consider different fluidic properties such as density, viscosity, surface tension, diffusion coefficient, electrical conductivity, and thermal conductivity of each fluid. Also, channels of the proposed 3D fluidic network
model are initially assumed to be filled with air similar to wet-lab experiments.

\setlength{\tabcolsep}{2.2pt}
\begin{table}[!h]
\tiny 
\caption{ Computation platform} 
\label{tb1:computation_platform}
\begin{center}
 \begin{tabular}{||>{\small}p{2.2cm}| >{\small}p{5.6cm}||}
 \hline
 CPU & Intel Core $i7-3770 $CPU @ $3.40$ GHz, $4$ cores \\
 \hline 
 Operating System & Windows 10  \\
 \hline
 &\\ 
  Software & COMSOL Multiphysics 5.2  \T\\
 \hline
 Model Dimension & 3D \T\\
 \hline
 \begin{tabular}{l}
 Materials \\(incompresible\\ Newtonian\\ Fluid) 
 \end{tabular} 
 & \begin{tabular}{l l}
 \multicolumn{2}{l}{Water used as buffer}\\
\hline 
 Sample & Diffusion co-efficient of \\& sample in water $[m^2/sec]$ \\ 
 \hline
 Ethanol & $1.24\times10^{-9}$\\
 Glycerol & $0.94\times10^{-9}$\\
 \hline
 \end{tabular} \B\\
 &\\ 
 \hline
  Used Physics & Laminar flow (SPF: Single Phase Flow) \T\\ 
  & Transport of diluted species (TDS)\T\\
 \hline
  Meshing & Predefined element size at extra fine while calibrating fluid dynamics \T\\
 \hline
 Study & Time dependent \T\\&(Time range: 0 sec - 200 sec; time step 1 sec) \\
 \hline 
 Solver & Direct PARDISO solver\T\\
 \hline 
\end{tabular}
\end{center}
\end{table}
\renewcommand{\arraystretch}{1}

Throughout all experiments, we consider the computational platform as summarized in Table \ref{tb1:computation_platform}. We have used the  following two physics of COMSOL Multiphysics software to study the physical profile of fluids flowing through the channels.
\begin{itemize}
\item Single-phase laminar flow (SPF): It is utilized for studying fluid velocity, fluid pressure, etc.
\item  Transport of diluted species (TDS): It is used for studying the concentration profile.
\end{itemize}


\input{tables/NumberOfElementsin3Series}

For analyzing the dynamics of the fluids, \emph{SPF} uses the  Navier-Stokes Equation that captures the conservation of momentum of fluids~\cite{Re_NS}. Equation~\ref{eq:NSlaw} represents the 
Navier-Stokes equation for incompressible Newtonian fluids:
\begin{equation}
\label{eq:NSlaw}
\rho (\frac{\delta u}{\delta t} + u\cdot \bigtriangledown u) = -\bigtriangledown p + \bigtriangledown \cdot ( \mu(\bigtriangledown u + (\bigtriangledown u)^T) - \frac{2}{3} \mu (\bigtriangledown \cdot u) I) + F
\end{equation}
where velocity, pressure, density, and dynamic viscosity of fluid are represented by \textit{u}, \textit{p}, $\rho$, and $\mu$, respectively. In Equation \ref{eq:NSlaw}, inertial forces are represented by the left-hand-side term; the right-hand-side term represents the total effect of pressure force ($-\bigtriangledown p$), viscous force ($\bigtriangledown \cdot ( \mu(\bigtriangledown u + (\bigtriangledown u)^T) - \frac{2}{3} \mu (\bigtriangledown \cdot u) I)$), and external forces ($F$). Equation \ref{eq:Continuity} represents the continuity which follows from the conservation of mass of fluid. \begin{equation} \label{eq:Continuity}
\frac{\delta \rho}{\delta t} + \bigtriangledown \cdot (\rho u) = 0
\end{equation} This continuity equation \cite{Re_NS} helps to solve these forces. Fick's law \cite{Crank} is used to compute the relative concentration of the sample. Also, the net transport is computable in the proposed fluidic network model. We have successfully used \emph{TDS} physics along with the transport equation (Equation \ref{eq:TDS}):
\begin{equation} \label{eq:TDS}
\bigtriangledown \cdot ( -D_i \cdot \bigtriangledown c_i) + u \cdot \bigtriangledown c_i = R_i
\end{equation}
where $D_i$ represents the diffusion coefficient, $c_i$ is the concentration factor for fluid $i$, and $R_i$ stands for the rate of reaction of the
fluids.

%% file: tables/NumberOfElementsin3Series.tex
\setlength{\tabcolsep}{2.2pt}
\renewcommand{\arraystretch}{1.2}
\begin{table}[b]
	\caption{Comparison among the number of elements in three series}
	\begin{center}
		\begin{tabular}{||>{\small\centering}p{1.15cm}|>{\small}p{2cm}|>{\small}p{2cm}|>{\small}p{2.5cm}||}
			\hline
			
			Accuracy\em{ (n) }
			& {\# elements in Farey sequence of order \mbox{$m$ (= $2^{n+1}-3$)}}
			& {\# elements in {\em FSD} sequence of order $2^n$ \mbox{(3$\times$ ($2^n-1$))}}
			& {\# elements in binarized sequence of \mbox{order} $n$ ($2^n+1$)} \\
			\hline
			\hline
			2
			& $F_5$ = 11 &$FSD_4$ = 9& $BS_2$ = 5\\
			3
			&$F_{13}$ = 59 & $FSD_8$ = 21&$BS_3$ = 9\\
			4
			&$F_{29}$ = 271 &$FSD_{16}$ = 45&$BS_{4}$ = 17\\
			5
			&$F_{61}$ = 1163 & $FSD_{32}$ = 93&$BS_{5}$ = 33\\
			6
			&$F_{125}$ = 4797 & $FSD_{64}$ = 189&$BS_{6}$ = 65\\
			\hline
		\end{tabular}
		\label{tbl:WorkDone}
	\end{center}
\end{table}


%% file: Text/Results.tex
\begin{table}[t]
  \centering
  \caption{Target \emph{CF}s For Different Dilution Gradients \cite{Sukanta_Grad}}
    \begin{tabular}{|l|c|p{9em}|c|}
    \cline{2-4}\multicolumn{1}{c|}{} & \multicolumn{1}{m{7.215em}|}{Dilution System} & \multicolumn{1}{m{10.215em}|}{~~~~~~Target-{\em CFs}} & \multicolumn{1}{m{4.215em}|}{Accuracy level ($n$)} \\ \hline
    \multicolumn{1}{|c|}{\multirow{6}{25pt}{Special test cases}} & Gaussian & \{$\frac{7}{128}, \frac{24}{128}, \frac{61}{128}, \frac{107}{128}$\} & \multirow{6}{*}{7} \\\hhline{~--~}
          & sin $x$ & \{$\frac{40}{128}, \frac{76}{128}, \frac{104}{128}, \frac{122}{128}$\} &  \\\hhline{~--~}
         & $x^\frac{1}{2}$ & \{$\frac{58}{128}, \frac{81}{128}, \frac{99}{128}, \frac{114}{128}$\} &  \\\hhline{~--~}
          & $x^2$  & \{$\frac{5}{128}, \frac{20}{128}, \frac{46}{128}, \frac{82}{128}$\} &  \\\hhline{~--~}
         & Linear & \{$\frac{26}{128}, \frac{51}{128}, \frac{77}{128}, \frac{102}{128}$\} &  \\\hhline{~--~}
          & 2-fold Log & \{$\frac{8}{128}, \frac{16}{128}, \frac{32}{128}, \frac{64}{128}$\} &  \\ \hline
    \multicolumn{1}{|c|}{\multirow{3}{25pt}{Real-life test cases}} & \multicolumn{1}{p{7.215em}|}{Bradford protein assay} & \multicolumn{1}{p{10.215em}|}{\{$\frac{10}{256}, \frac{20}{256}, \frac{30}{256},$ $\frac{40}{256}, \frac{50}{256}$\}} & \multirow{3}{*}{8} \T\B\\ \hhline{~--~}
          & \multicolumn{1}{p{7.215em}|}{Sucrose gradient analysis} & \{$\frac{10}{256}, \frac{15}{256}, \frac{20}{256},$ $\frac{25}{256}, \frac{30}{256}, \frac{35}{256}, \frac{40}{256}$\} &  \\\hhline{~--~}
         & \multicolumn{1}{p{7.215em}|}{Harmonic gradient analysis} & \{$\frac{128}{256}, \frac{85}{256}, \frac{64}{256}, \frac{51}{256}, $ $\frac{42}{256}, \frac{36}{256}, \frac{32}{256}, \frac{28}{256}$\} &  \\ \hline
    \end{tabular}%
  \label{tab:addlabel}%
\end{table}%

We performed experiments for comparing the accuracy
of the \emph{FSD} based discretization and binarized discretization (\emph{BS}) sequence, and observed that for an accuracy value $n$, \emph{FSD} provides denser discretization of the concentration range $[0, 1]$. For example, for accuracy level 6,  binarization based discretization splits the interval $[0, 1]$ into $64$ segments  as shown in Table \ref{tbl:WorkDone};  whereas in \emph{FSD}, it is split into $188$ intervals. Therefore, it is evident that in the proposed \emph{FSD} framework,
\begin{figure}[h]
	\centering
	\begin{tabular}{c }
		\frame{\includegraphics[width=6.8cm,keepaspectratio]{pics/21_scatter_plot_BDPvsBS.pdf}}\\
		(a) Scatter-plot (error \textit{vs} target-\textit{CF}) of 200 experimental data. \\
		\frame{\includegraphics[width=6.8cm,keepaspectratio]{pics/22_Total_AvgError.pdf}}\\
		\multicolumn{1}{p{8.5cm}}{(b) {Average differential cost of sample and buffer is shown along the $Y$-axis with respect to theoretical requirement.}}\\
	\end{tabular}
	\caption{Comparison between {\em FSD} and {\em BS}-based methods \cite{MinMix,Banerjee2017} for accuracy 9.}
	\label{fig:ScatterPlot}
\end{figure} 
 a given target-\emph{CF} is approximately more accurately compared to the conventional 
binary discretization method. 

We further performed experiments on $200$ test-cases covering several random target-\emph{CF}s, some real-life test-cases such as \emph{Bradford Protein Assay},\emph{ Sucrose Gradient Analysis}, and \emph{Harmonic Gradient Analysis}, and on some special dilution gradients e.g., \emph{Gaussian}, \emph{sin} $x$, $\surd x$, $x^2$, \emph{Linear}, \emph{2-fold Log}~\cite{Sukanta_Grad} (shown in Table~\ref{tab:addlabel}). We notice that {\em CF}-errors generated by the proposed method become much smaller than those in other methods~\cite{Banerjee2017,MinMix}. This is also reflected from Fig. \ref{fig:ScatterPlot} (a) where {\em CF}-errors generated by the proposed method lies closer to the $Y$-axis. 
We also define {\it differential cost} of a reactant as follows: the absolute difference between reactant requirement as mandated by user's specified target-$CF$ and that actually needed when its nearest approximation on the discretized concentration range is used in computation. For both $FSD$- and $BS$-based methods, we record the differential cost of sample and buffer averaged over all discretized values of target-$CF$s (Fig.~\ref{fig:ScatterPlot}(b)). The results clearly show that $FSD$ outperforms $BS$-based methods~\cite{MinMix,Banerjee2017}.

\begin{figure}[h]
    \centering
    \frame{\includegraphics[width=7cm,height=5.0cm]{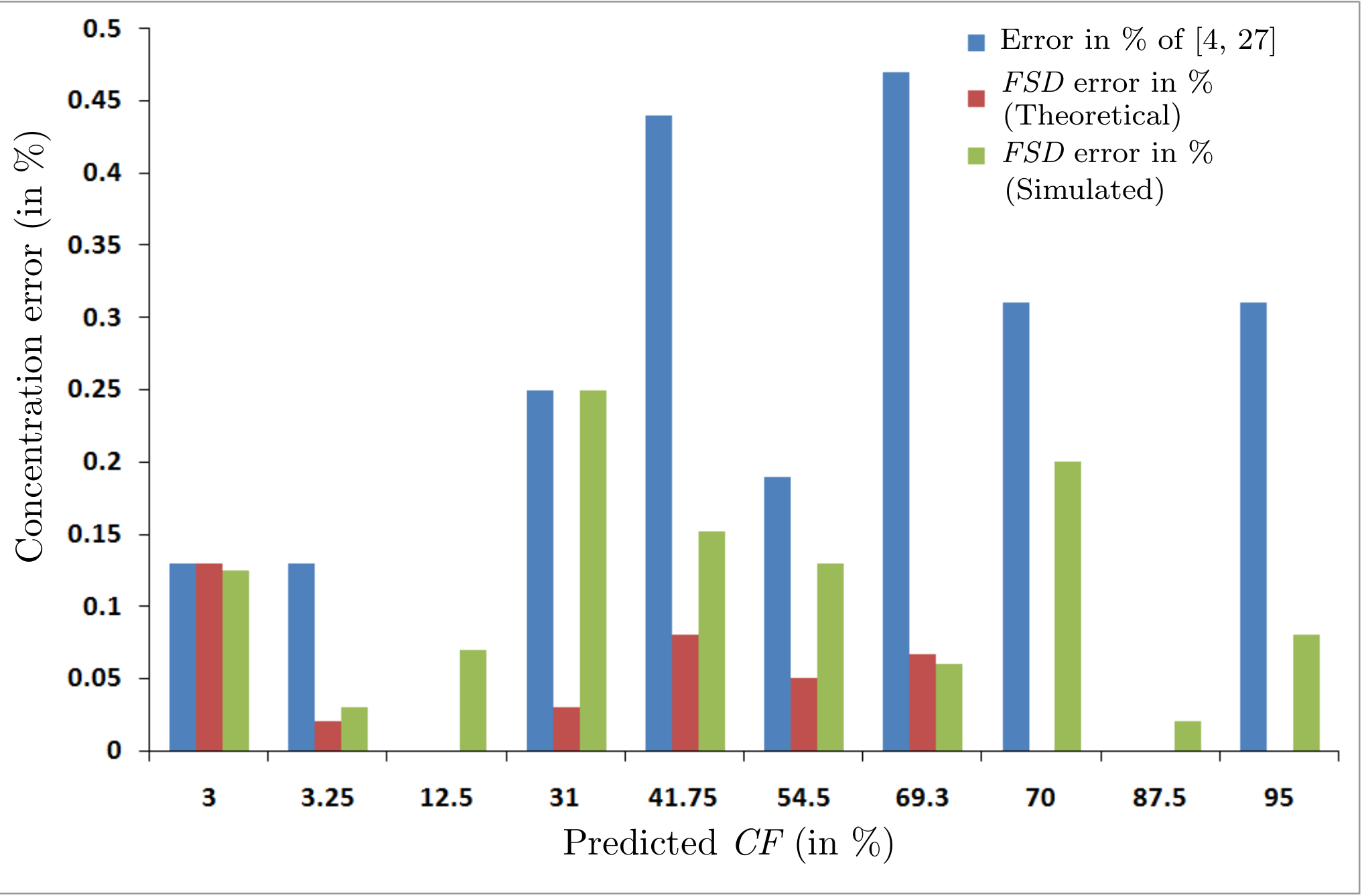}}
    \caption{Plot showing theoretical and experimental absolute error values in concentration factors for several test-cases.}
    \label{fig:ErrorGraph}
\end{figure}

We also compare the theoretical {\em CF}-errors of randomly generated synthetic test-cases \{3\%, 3.25\%, 12.5\%, 31\%, 41.75\%, 54.5\%, 69.3\%, 70\%, 87.5\%, 95\%\} by {\em FSD} method and \emph{BS}-based method. As before, we observe \emph{FSD} provides better accuracy then \emph{BS}-based method. We run
COMSOL simulation on these  synthetic test-cases and found that experimentally observed $CF$-errors favorably match with theoretically estimated {\em CF}-errors of {\em FSD} (see Fig.~\ref{fig:ErrorGraph}).


\begin{figure}[h]
    \centering
    {\includegraphics[width=7cm,height=5.2cm]{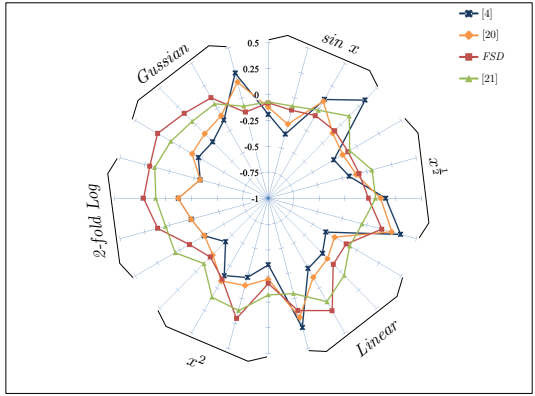}}
    \caption{Radar chart representation of errors in \emph{tCF}s of some gradient test cases for the proposed
    method and three other methods with  accuracy 7.}
    \label{fig:RadarChart}
\end{figure}

{\color{black}
Moreover, we compare the performance ({\em w.r.t.} accuracy) of {\em FSD}  with prior approaches~\cite{Banerjee2017,BanerjeeVLSID2019,BanerjeeCBT}. To accomplish this task, we further run COMSOL simulation on several concentration-gradient  test-cases for accuracy 7 (as listed in Table \ref{tab:addlabel}). A radar-chart is shown in Fig. \ref{fig:RadarChart}, where the radial axis reflects the errors in observed \emph{CF}-values.  This chart demonstrates that the proposed \emph{FSD} network efficiently generates the target \emph{CF}s with much  less error compared to existing methods \cite{Banerjee2017,BanerjeeVLSID2019,BanerjeeCBT}.}

\begin{figure}[!h]
    \centering
    {\includegraphics[width=7cm,keepaspectratio]{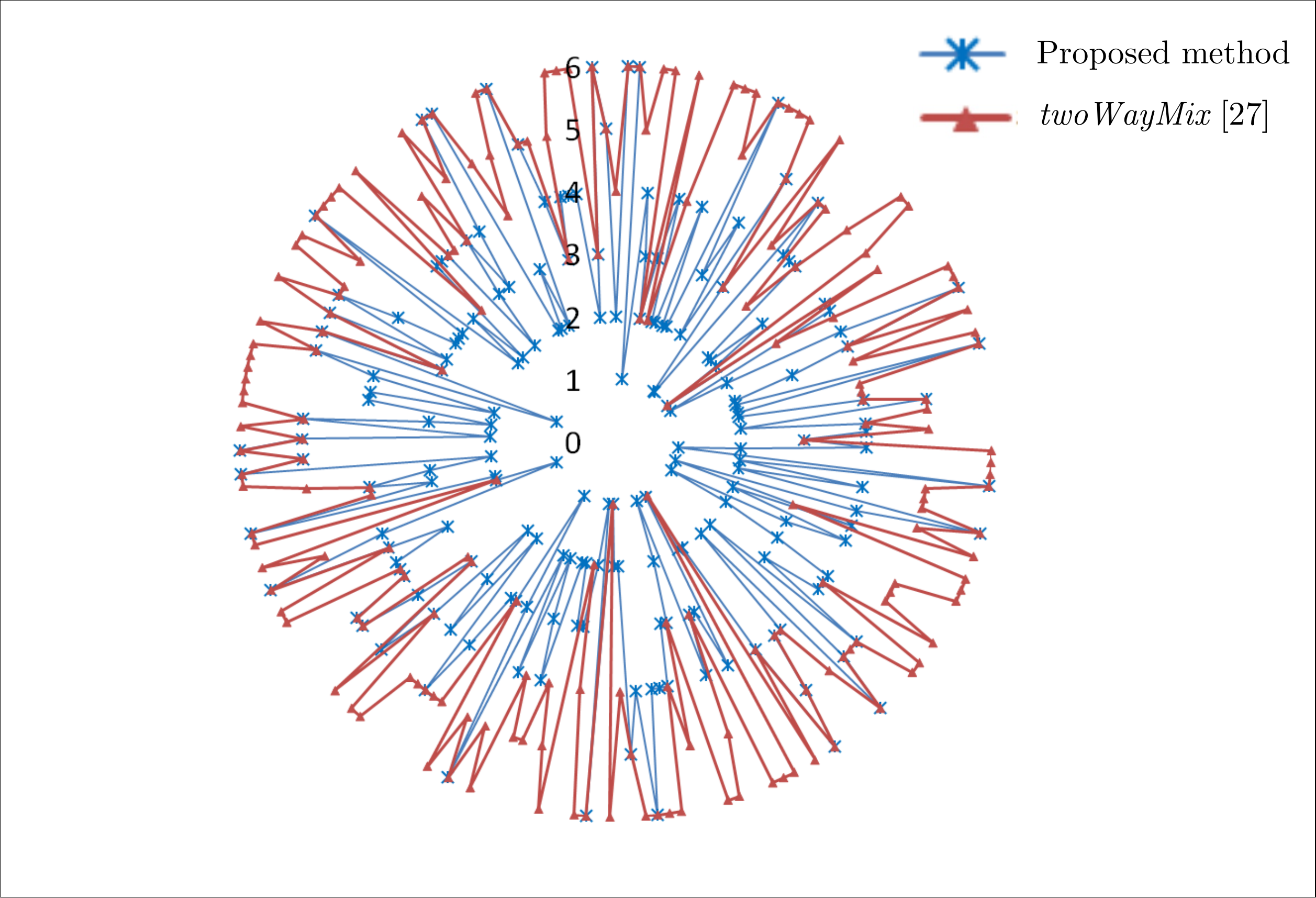}}
    \caption{Comparison with respect to number of mixing steps required to generate different target-\emph{CF}s by \cite{MinMix} and the proposed method.}
    \label{fig:StepsRadar}
\end{figure}

\begin{figure}[!h]
    \centering
    \includegraphics[width=9cm,height=4.3cm]{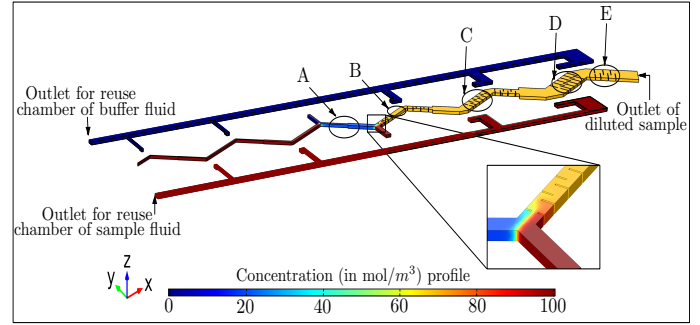}
    \caption{Flow dynamics in the mixing channel.}
    \label{fig:flowDynamics}
\end{figure}

\begin{figure}[!h]
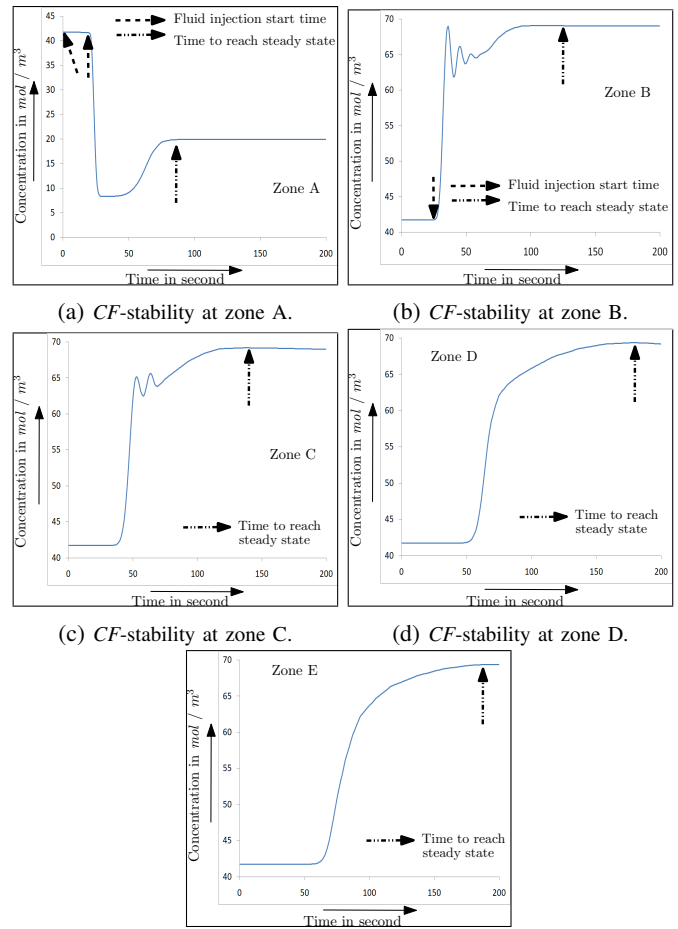

	\centering
	\begin{tabular}{p{2.15cm} c c c  }
		\multicolumn{2}{c}{\frame{\includegraphics[width=4.3cm,keepaspectratio]{pics/24_CF_timeSlice_A.pdf}}} & \multicolumn{2}{c}{\frame{\includegraphics[width=4.3cm,keepaspectratio]{pics/25_CF_timeSlice_B.pdf}}}\\
		\multicolumn{2}{c}{(a) {\footnotesize{\em CF}-stability at zone A.}} & \multicolumn{2}{c}{(b) {\footnotesize{\em CF}-stability at zone B.}}\\
		\multicolumn{2}{c}{\frame{\includegraphics[width=4.3cm,keepaspectratio]{pics/26_CF_timeSlice_C.pdf}}} & \multicolumn{2}{c}{\frame{\includegraphics[width=4.3cm,keepaspectratio]{pics/27_CF_timeSlice_D.pdf}}}\\
		\multicolumn{2}{c}{(c) {\footnotesize{\em CF}-stability at zone C.}}&\multicolumn{2}{c}{(d) {\footnotesize{\em CF}-stability at zone D.}}\\
		\multicolumn{1}{c}{}&\multicolumn{2}{c}{\frame{\includegraphics[width=4.3cm,keepaspectratio]{pics/28_CF_timeSlice_E.pdf}}}&\\
		&\multicolumn{2}{c}{(e) {\footnotesize{\em CF}-stability at zone E.}}&
	\end{tabular}
    \caption{ (a)-(e) Convergence of concentration factor with time at different zones (Fig. \ref{fig:flowDynamics}) of the fluidic network for \emph{tCF} = 69.3\%.}
    \label{fig:TimeSlice}
\end{figure}

We perform additional experiments on 200 randomly generated target-\emph{CF}s to compare the number of mixing 
steps required by the proposed method and {\em twoWayMix}~\cite{MinMix}.
We show the results in 
Fig. \ref{fig:StepsRadar} from which it can be noticed that the {\em FSD} method requires fewer mixing steps in most of the target-\emph{CF}s compared to {\em twoWayMix}~\cite{MinMix}. Also, to investigate the concentration
stability in different zones (zones A-E in Fig.~\ref{fig:flowDynamics}) of the proposed layout, we captured the flow-dynamics of the mixing channels while producing the target-\emph{CF} = $\frac{9}{13}$ using COMSOL Multiphysics Software. We show the convergence of \emph{CF}s with time at different zones (A-E) in Fig.~\ref{fig:TimeSlice}. We observed that the concentration stability 
in Zone A occurs approximately after 80 seconds. Whereas, in Zone B and Zone C, stability is achieved approximately after 125 seconds and 140 seconds, respectively. Zone D and Zone E reach the steady-state after 200 seconds, following which flow with the desired target concentration factor is steadily observed at the outlet.

%% file: Text/Conclusions.tex
In this paper, we have presented a novel architecture of free-flowing biochip that can be used to perform dilution of sample fluids accurately. This work mainly has two contributions: (i) in contrast to traditional binary discretization of the concentration interval $[0, 1]$, we first propose a new technique for approximating a target-\emph{CF} based on Farey-sequence (\emph{FS}) arithmetic, and (ii) secondly, we present the detailed design of a befitting microfluidic architecture that mimics the proposed \emph{FS}-based \emph{CF}-approximation rule and subsequently leads to an efficient implementation of the diluter. The network is free-flowing in nature, i.e., it does not need any control valve for effecting conditional navigation of fluids through the interior labyrinth of channels. Flow-controllers are required only at the inlet ports to enable the fluid to enter either into the mixing channel or towards re-use branches. The state of each controller can be set {\em a priori} (off-line) depending on the given target-\emph{CF}. Additionally, no differential pressure/velocity control is needed for injecting input-fluids into the network, and thus sustenance of constant-pressure would suffice. Being free-flowing in nature, no controlled splitting or metering of sample/buffer units is required during the execution of the dilution assay. The network supports programmability in the sense that dilution of fluid targeting any concentration factor within the range $[0, 1]$ can be achieved with no more than $1\%$ error using this chip. We simulated the network architecture using COMSOL Multiphysics Software and studied its performance for various test-cases. We observe that error in target-\emph{CF}, convergence time, and sample/buffer cost for the \emph{FSD}-based method compare favorably against those achievable with other competing flow-based systems. The proposed fluidic network can thus provide a potential platform for automating sample preparation.